# A New Kind of Beauty Out of the Underlying Scaling of Geographic Space

Bin Jiang[1] and Daniel Sui[2]

[1]Department of Technology and Built Environment, Division of Geomatics
University of Gävle, SE-801 76 Gävle, Sweden
Email: bin.jiang@hig.se

[2]Department of Geography, and Center for Urban and Regional Analysis
The Ohio State University, Columbus, OH 43210, U.S.A
Email: sui.10@osu.edu

*(Draft: March 2012, revision: December 2012, March, May 2013)*

**Abstract**
Geographic space demonstrates scaling or hierarchy, implying that there are far more small things than large ones. The scaling pattern of geographic space, if visualized properly (i.e., based on head/tail breaks), can evoke a sense of beauty. This is our central argument. This beauty is a new type of aesthetic at a deeper structural level, and differs in essence from an intuitive sense of harmony, perceived in terms of color, shape, texture, and ratio. This new kind of beauty was initially defined and discovered by Christopher Alexander, and promoted in his master work *The Nature of Order*. To paraphrase Mandelbrot, this is the beauty for the sake of science rather than for art's sake or for the sake of commerce. Throughout the paper, we attempt to argue and illustrate that the scaling of geographic space possesses this new kind of beauty, which has a positive impact on human well-being. The paper further draws upon the previous work of Nikos Salingaros and Richard Taylor on the beauty in architecture and arts to support our argument.

## 1. Introduction

A fundamental aspect of geography as a mode of scientific inquiry is to understand forms and processes of geographic systems involving both natural and built environments. With the increasing availability of geographic information, a universal form or pattern across different scales called scaling has received growing attention in geography and related disciplines (Puman 2006). Scaling has been extensively studied in economics (Eeckhout 2004), architecture and urban planning (Salinaros 2005, 2006), music (Manaris et al. 2005), computer science (Mitzenmacher 2004), physics (Newman 2005), biology (Bonner 2006), and mathematics (Mandelbrot 1982). Scaling is also referred to as scale-free, scale invariance, fractal, hierarchy, and nonlinearity. The underlying mathematics of scaling patterns in different contexts, due to their differing origins, has also been given different names such as Zipf's law, Pareto distribution, and power law distributions, all of which have been widely recognized as the 80/20 principle or long tail theory in the popular press and business literature (Koch 1998, Anderson 2006). This paper uses "scaling" because it is widely used and literally related to the term "scale." In this paper, scale simply refers to size rather than map scale, resolution, or phenomena scale as used in the geography literature. Unfortunately, geographers took the term scale from other disciplines for granted, often without paying due attention to the delicacies of its true meanings (e.g., Goodchild and Mark 1987, Lam and Quattrochi 1992, Batty and Longley 1994). It is important to note that scale is closely related to scaling, i.e., different scales ranging from the smallest to the largest form a scaling hierarchy. The scaling hierarchy has some profound implications for understanding human mobility (Jiang 2009, Jiang et al. 2009), cartographic mapping (Jiang 2012a, Jiang et al. 2012), and cognitive mapping (Jiang 2012b). These implications can be extended to human emotions about the aesthetic impacts of the scaling patterns of geographic space.

This paper argues that the beauty of geographic space arises from underlying scaling; that scaling, if visualized properly (based on head/tail breaks, c.f., Section 4), can evoke a sense of beauty. This beauty differs fundamentally from conventional wisdom about aesthetics, which is essentially based on two views. The first view regards beauty as spontaneous and intuitive, without need for explanation - *"beauty is in the eye of the beholder,"* as the cliché goes. There are multiple parameters for assessing aesthetic qualities such as color, shape, texture, and ratio, but it is difficult, if not entirely impossible, to quantify these for an artistic product or a scientifically discovered pattern. In contrast to the first view, the second treats beauty as a quality that can be scientifically assessed based on empirical aesthetics (Birkhoff 1933). For example, the golden ratio $\Phi = 1.618$ is widely recognized as a measure for various beautiful shapes, such as animal skeletons, plant shapes, and human faces. It is also found in many historical buildings such as the Pantheon (Livio 2002), although Salingaros (2012) has recently challenged this long-standing view. These two views constitute a major part of the conventional understanding of aesthetics. The sense of beauty evoked by scaling is founded on a newly recognized kind of aesthetics at the profound level of structure (Alexander 2001). To paraphrase Mandelbrot (1989), this is beauty for the sake of science rather than for the sake of arts or commerce. This new kind of aesthetic experience may not be easily perceived for an untrained mind, and often requires effort and experience in order to discern it. In fact, the scaling pattern, at first glance, may appear to be arbitrary, irregular, random, or chaotic.

To further elaborate on this new aesthetic, an analogy of perceiving the human body as beautiful may be useful. The human body is perceived as beautiful due to a symmetrical structure and well-balanced components that can be characterized by the golden ratio. It is the appearance or surface structure that makes things beautiful. However, the beauty of the human body can be assessed at a deeper structural level. A human body consists of many organs, each of which consists of many tissues or types of tissues. The tissues are composed of numerous cells. These cells are composed of an immense number of molecules, and those molecules are made up of uncountable atoms. The different scales of structure, ranging from the organ down to the atom, constitute the whole human body. This forms a scaling hierarchy or whole. This hierarchy or whole makes the human body highly intelligent, alive, harmonious, and beautiful. Geographic spaces possess the same kind of scaling structure, in which there are far more small geographic features than large ones. It is the underlying scaling of geographic space that triggers a sense of beauty (c.f., Section 4, and 5). This is the central argument we want to make in this paper.

The contribution of this paper is three-fold: (1) we associate head/tail breaks and its derived scaling hierarchy to the theory of centers, and in particular three of the 15 structural properties (Alexander 2001); (2) we demonstrate that the scaling of geographic space as a whole triggers a sense of beauty in the deep psyche, and this is supported by the theory of centers; and (3) we draw upon related evidence in architecture and arts to support our central argument that beauty arises out of the underlying scaling of geographic space. By doing this, we link scaling in general - or scaling of geographic space in particular - to Alexander's theory of centers to assess the new kind of beauty for geographic space.

The remainder of this paper is organized as follows. Section 2 presents two basic concepts of scales and scaling, and their relationship in order to better understand the new kind of aesthetic. Section 3 briefly introduces the theory of centers and related concepts such as centers, wholeness, life and the 15 structural properties. Section 4 introduces head/tail breaks with a workable example based on the Koch snowflake structure, and associates it with the theory of centers. Section 5 presents three case studies to illustrate how beauty arises from the underlying scaling of geographic space based on the theory of centers. Section 6 further draws on related evidence in the fields of architecture and arts to support the argument that scaling patterns are capable of evoking a sense of beauty. The last section contains concluding remarks and discussions of some broader implications of Alexander's theory for future geographic research.

## 2. The concepts of scales and scaling

Scales, scaling, and their relationship are fundamental to the argument we offer, so it is of paramount importance to clarify these concepts with a simple example. The famous Koch snowflake is composed of equilateral triangles of three scales (or sizes of 1, 1/3, and 1/9) (Figure 1). The numbers of the triangles for the three scales are respectively 1, 6, and 18. The medium-sized triangles are one-third of the largest triangle, and the smallest triangles are again one-third of the medium-sized ones. In other words, the smaller triangles are similar by a similarity ratio of 1/3. There are far more small triangles than large ones, or equivalently, far more small scales than large ones. This is a scaling pattern across the three different scales. Many readers may refer to this as a fractal pattern, which is indeed the case, for "fractal" and "scaling" are interchangeably used in this paper. The Koch snowflake exhibits an exact self-similarity, while patterns observed in nature and society may be more likely to bear statistical self-similarity, such as coastlines or other geographic features (Mandelbrot 1967, 1982).

The surface of the earth or geographic space involves many individual scales that form a scaling hierarchy. The smallest scale that human beings can perceive is probably at the scale of meters, on the order of $10^0$, while the largest scale is at the order of $10^7$, as large as the planet. Most of the time, the scale we are usually concerned with is at the city or regional level, typically around $10^4$. This range ($10^0$–$10^4$) is in terms of the physical extent of cities. The scales of geographic spaces can be measured with other parameters, such as city sizes by population. The scale range for cities in terms of population in the United States can be $10^3$–$10^7$, with the ratio about $10^4$. In an architectural space, the smallest scale can go down to 3 mm ($10^{-3}$) at the ornament level, while the largest scale can be up to $10^1$ or $10^2$.

There are a few intermediate scales between the smallest and the largest. Importantly, the numerous structures or substructures of different scales form a scaling hierarchy. There are far more small things than large ones. For example, there are far more small cities than large cities (Zipf 1949), far more low buildings than high buildings (Batty et al. 2009), far more short streets than long streets (Carvalho and Penn 2004) or far more less-connected streets than highly-connected streets (Jiang 2009), and far more small street blocks than large street blocks (Jiang and Liu 2012). This list can go on and on to include scaling patterns of architectural space. In fact, Salingaros (2006) has established a theory of architecture based on this kind of scaling hierarchy.

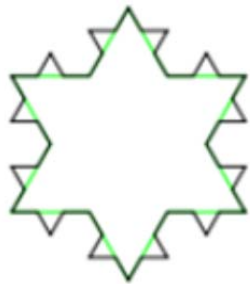

Figure 1: The scaling pattern of a Koch snowflake across three different scales

Scaling also applies to living organisms. Taking the same example of a human body mentioned earlier, there are a huge number of atoms (at the order of $10^{27}$), yet only one ($10^0$) is the largest – the human body itself. Between the smallest (atoms) and the largest (body), there are a few intermediate scales: molecules, cells, tissues, and organs. Physicist Erwin Schrödinger (1944) raised an intriguing question: Why are atoms so small? A typical answer is that human beings are so highly intelligent and complex that they need such small building blocks to organize the complexity. Alternatively, the small units are essential for the highly complex human body not to be affected by microscope fluctuation. A parallel question would ask why there are so many atoms. The answer is similar to the previous one: the numerous atoms help the human body achieve its high level of complexity. Different scales, ranging from the smallest to the largest, meet an inverse power law relationship, $p \cdot x^\alpha = C$, where $p$ is the number of units, $x$ is the scale, $a$ is the power law exponent, and $C$ is the constant (Salingaros 2005, Chapter 3). The power law is one of the mathematic formulas for characterizing scaling or scaling hierarchy.

## 3. A new kind of aesthetic at a deep structural level
According to Alexander (1993, 2001), the theory of centers is applicable to both organic and inorganic phenomena, ranging in scale from $10^{-15}$ to $10^{26}$ meters, covering almost the entire universe. The theory notably abandons a mechanistic twentieth century worldview and adopts instead a brilliant, new organic worldview; in this sense, it represents a monumental shift in scientific and artistic thinking. The theory helps bridge science and the arts by providing a set of scientific tools to objectively judge beauty. The fundamental thesis of Alexander's master work *The Nature of Order* (Alexander 2001) is that the order in both nature and what we build is essentially the same - the harmonious order that fills and touches human beings with an acute sense of wholeness/holism.

The theory provides a set of concepts by which to judge the beauty of structure or patterns. This beauty is founded in a new kind of aesthetics that is essentially different from one's intuitive sense of beauty, evoked by pretty colors or joyful design. The basic concepts of the theory are wholeness and centers. Any well-designed structure is a whole. The whole consists of many interconnected and mutually reinforced centers, so center is the basic building block of a structure. For example, a human face is a whole, consisting of many centers, such as the eyes, nose, and mouth. Note that centers are found at different levels of scale. Eyebrows and eyelashes are centers at a finer scale. A city is a whole, consisting of many centers, such as a city hall, parks, and some major streets. The city hall is a whole, consisting of many centers such as windows, rooms, and floors. There are three points worth noting: 1) centers are defined at the different levels of scale; 2) centers are field-like entities, not necessarily having clear boundaries; and 3) centers are interconnected and influence each other to form a whole. It is important to note that centers can have either a positive or negative effect in forming a whole. Figure 2 illustrates an example showing a rectangle as a whole (Alexander 2001); the tiny dot outside the rectangle detracts from the whole, while the same dot inside reinforces the whole of the rectangle.

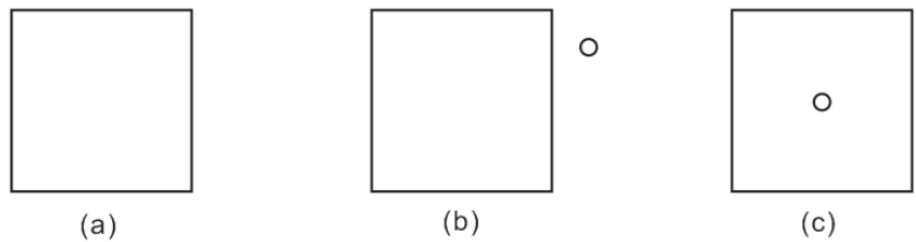

(a)          (b)          (c)

Figure 2: Positive or negative effects of a center on a structure
Note: The rectangle as a center (a), the center detracted by the dot (b), and reinforced by the dot (c)

Beginning with the basic question of how to make beautiful buildings, and based on his classic work of pattern language, Alexander (2001) distilled 15 geometric properties (Table 1) used to create beautiful patterns, to judge the beauty of a structure, and to guide a design in general. To judge its beauty is simply to examine how something is able to achieve wholeness through the interaction of numerous centers. The most beautiful things are those achieving the highest degree of wholeness; the more dense the appearance of the 15 properties, the more beautiful the underlying structure. Wholeness is what resonates deeply with human experience; the perception of such a wholeness or underlying structure gives rise to the human experience of beauty. The quality of wholeness is definite, tangible, and objective, rather than a matter of preference or taste, so the beauty emerging from deep structure is an objective judgment rather than a subjective preference (Alexander 2001).

Table 1: The 15 properties as the glue that binds centers together

| | | |
|---|---|---|
| Levels of scale | Good shape | Roughness |
| Strong centers | Local symmetries | Echoes |
| Thick boundaries | Deep interlock and ambiguity | The void |
| Alternating repetition | Contrast | Simplicity and inner calm |
| Positive space | Gradients | Not separateness |

Figure 3 illustrates three of the 15 geometric properties with the snowflake structure shown in Figure 1. The three geometric properties illustrated here are the most important, and are relatively easy to identify with different structures or patterns. The first property is *levels of scale*, or the definition of centers at a range of sizes. The snowflake contains three levels of scale: 1, 1/3, and 1/9, as mentioned earlier. But the structure can be refined iteratively to smaller or finer scales. The second property is *strong centers*. The figure indicates that the strongest center is supported by six centers, which are themselves further supported by the 18 tiny centers (Figure 3). Centers are often formed and strengthened by *boundaries*, which are the third property. The triangles at different scales are the boundaries of the centers. In Figure 3, the low-level centers support the larger ones in a recursive manner, while the largest one receives the most support. Note that it is not easy to identify centers or the 15 geometric properties in general, with various structures in nature or in artifacts. Appreciation of this type of beauty is not spontaneous, but is developed through experience and devoted efforts.

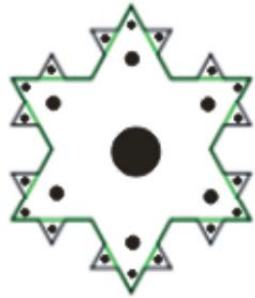

Figure 3: Illustration of levels of scale, strong centers, and boundaries

A structure with a high degree of wholeness is said to have life in it. This concept of life goes beyond biological life to include both animate and inanimate things as long as they possess the scaling structure. In terms of Alexander (2001), *"[life] is a quality which inheres in space itself, and applies to every brick, every stone, every person, every physical structure of any kind at all, that appears in space."* It is life that links to humans' deep feelings, and has a positive emotional impact. It was found that patterns with a living structure have greater aesthetic impact on human beings (Alexander 1993, 2001). It is in this sense that the notion of beauty or living structure is objective rather than subjective. Alexander (2001) has called for new mathematics to quantify the degree of beauty or life. Beauty is not just in the eye of the beholder, but also in the deep level of structure. In the following sections, we will first introduce head/tail breaks (Jiang 2012b), and then examine how geographic patterns illustrated by head/tail breaks meet the three structural properties, being living structures. Thus, the patterns that reflect the scaling of geographic space tend to have high aesthetic impacts.

## 4. Head/tail breaks leading to classes being a whole

Head/tail breaks is a new classification scheme for data with a heavy-tailed distribution that follows power law, lognormal, or exponential function. The three functions are very different mathematically, but they share one common interesting property. If we rank data values that follow the three distributions from the largest to the smallest (or so called rank-size plot), all the values can be put into two imbalanced parts: a minority of values above the mean as the head, and a majority of values below the mean as the tail. This was formulated as the head/tail division rule (Jiang and Liu 2012). Interestingly, this head/tail division rule applies to the head iteratively, leading to the novel classification scheme of head/tail breaks (Jiang 2012b).

To illustrate, the mean size of the 25 triangles of the snowflake is $(1 \times 1 + \frac{1}{3} \times 6 + \frac{1}{9} \times 18)/25 = 0.2$. This first mean ($m_1$) puts the 25 triangles into two parts: 7 above the mean in the head (28%), and 18 below the mean in the tail (72%). For the 7 triangles in the head, the mean is $(1 \times 1 + \frac{1}{3} \times 6)/7 = 0.4$. This second mean ($m_2$) puts the 7 triangles into two parts again: 1 above the mean in the head (14%) and 6 below the mean in the tail (86%). Thus the two means lead to three classes: the first class between the minimum and $m_1$, the second class between $m_1$ and $m_2$, and the third class between $m_2$ and the maximum. These three classes correspond to the three scales as elaborated above.

The unique thing about the head/tail breaks lies in its simplicity. For a large dataset involving many classes or hierarchical levels, the head/tail breaking process continues until data values in the last head are no longer heavy-tailed distributed, or the values left in the last head are no longer a minority. For example, if from a previous iteration, 10 values are left in the head, and 5 out of the 10 are above the mean, then the breaking process should be terminated. This is how individual classes or hierarchical levels are derived.

Unlike other classification schemes, the head/tail breaks can automatically determine both the number of classes and class intervals, according to the very scaling property of the data. The following case studies adopt the head/tail breaks, rather than the commonly used natural breaks (Jenks 1963), to visualize scaling patterns. The natural breaks aims to minimize intra-class variance and in the meantime maximize inter-class variance. Although widely used, the natural breaks fails to capture the scaling structure of the data with a heavy-tailed distribution (Jiang 2012b). In fact, the head/tail breaks is more natural than the natural breaks for the data with a heavy-tailed distribution.

There are several reasons why the head/tail breaks is more natural than the natural breaks. First, the classification scheme is based on human binary thinking that tends to put things into two categories around an average value (e.g., large/small, long/short, rich/poor, and urban/rural). Second, any two consecutive classes from high to low reflect the head/tail division, and constitute an imbalanced contrast, with the high class distinguished from the low class forming Gestalt figure-ground perception. Unlike the famous Rubin's vase (Rubin 1921), the derived classes show less ambiguity. All classes together constitute a scaling hierarchy, thus capturing the scaling property of the data. Third, both the number of classes and class intervals are objectively determined, rather than subjectively imposed, such as the natural breaks. Finally, scaling is the essence of nature and society, in which there are far more small things than large ones. It is in this vein that the head/tail breaks is more natural than the natural breaks.

We claim that the classes derived by the head/tail breaks are able to form a whole, since the three structural properties are present with the resulting patterns of the classification. First, the derived classes can be considered to be centers, which are defined at the different *levels of scale*. Second, all heads constitute *strong centers* distinguished from the first class or first tail. The top most class is the strongest center. The breaks, or the different means, constitute the *boundaries* between classes. The presence of these three properties is mainly discussed at the data level rather than visual level. This will be shown and further discussed in the following case studies.

## 5. Beauty within scaling patterns of geographic space: Case studies
The following case studies illustrate that the scaling of geographic space can evoke a sense of beauty through perceiving the derived classes as a whole. The most beautiful patterns are those achieving the highest degree of wholeness. We compare three pairs of patterns, derived respectively by the head/tail breaks and the natural breaks, and elaborate on which one achieves the higher degree of wholeness. It is important to note that the classes derived using the head/tail breaks are able to form a whole based on the theory of centers, and the head/tail breaks method captures the underlying scaling better than the natural breaks.

### 5.1 Scaling pattern of U.S. cities
The first case study is to map the scaling pattern of U.S. cities by population. The data was taken from the 2012 U.S. Census and included 4,256 cities. The minimum city population (or size) was 10,005, while the maximum population was 7,322,564, so the ratio of the maximum to the minimum was about $10^3$, a clear indicator of scaling. The first mean value was 64,875, followed by second and third mean values of 315,257, and 1,178,085 in a recursive manner. This led to four classes or hierarchical levels shown in Figure 4a, and the rows of the head/tail breaks and related number of cities in Table 2. The numbers of classes, in a lower versus an upper class, met the imbalanced division of a majority versus a minority. This was not the case for the classes derived from the natural breaks (see the rows of the natural breaks and related number of cities in Table 2). The visual pattern in Figure 4a was full of surprises from one location to another, while the pattern in Figure 4b was very flat, indicating a monotonous structure underneath.

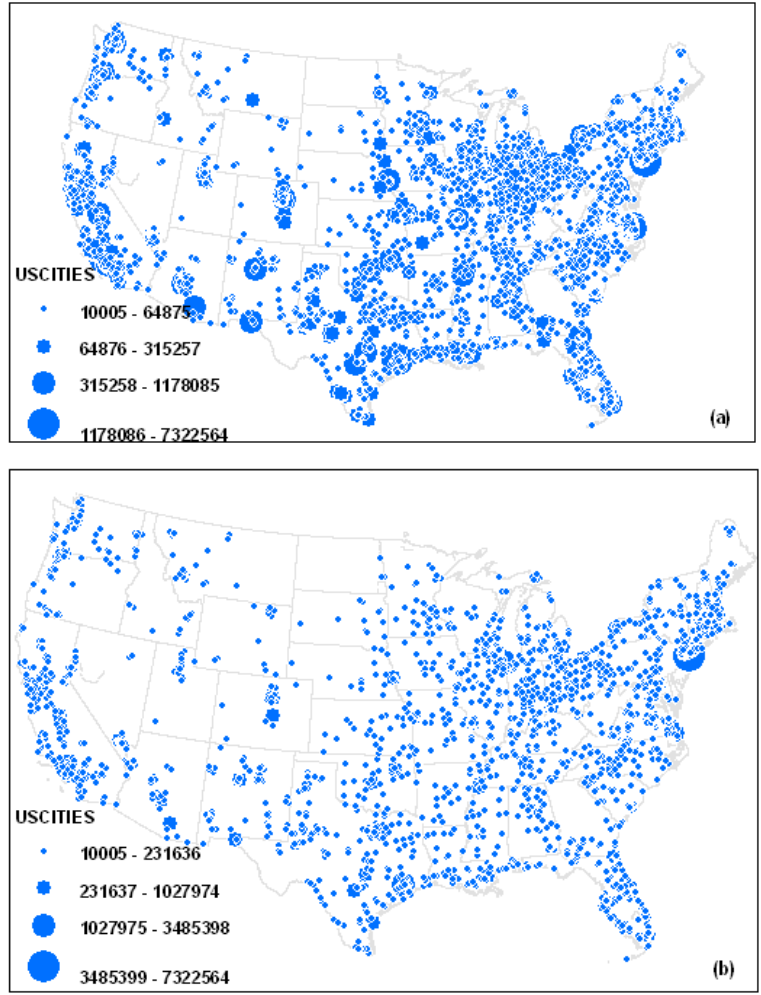

Figure 4: The scaling of U.S. city sizes (in terms of population) captured by the head/tail breaks (a) rather than the natural breaks (b)

Table 2: Four classes of U.S. cities in terms of population

| Classes | 1 | 2 | 3 | 4 |
|---|---|---|---|---|
| Head/tail breaks | 10,005–64,875 | 64,876–315,257 | 315,258–1,178,085 | 1,178,086–7,322,564 |
| Number of cities | 3,648 | 496 | 94 | 18 |
| Natural breaks | 10,005–231,636 | 231,637–1,027,974 | 1,027,975–3,485,398 | 3,485,399–7,322,564 |
| Number of cities | 4,118 | 120 | 10 | 8 |

Each dot, representing a city on the map, is a center, so there are more than 4,000 centers. They are at four levels of scale as seen in Figure 4a. The four classes are actually four centers perceived at a higher level. Among the 4,000 centers, the 18 largest cities are the strongest centers, reinforced by other surrounding centers. Note that the 18 largest cities cannot be broken up into two categories; otherwise, the head part would not be a minority (i.e., 8 out of 18 is greater than 40%). On the other hand, for the pattern using the natural breaks, centers look very flat, and many of them are not distinguished from their surroundings.

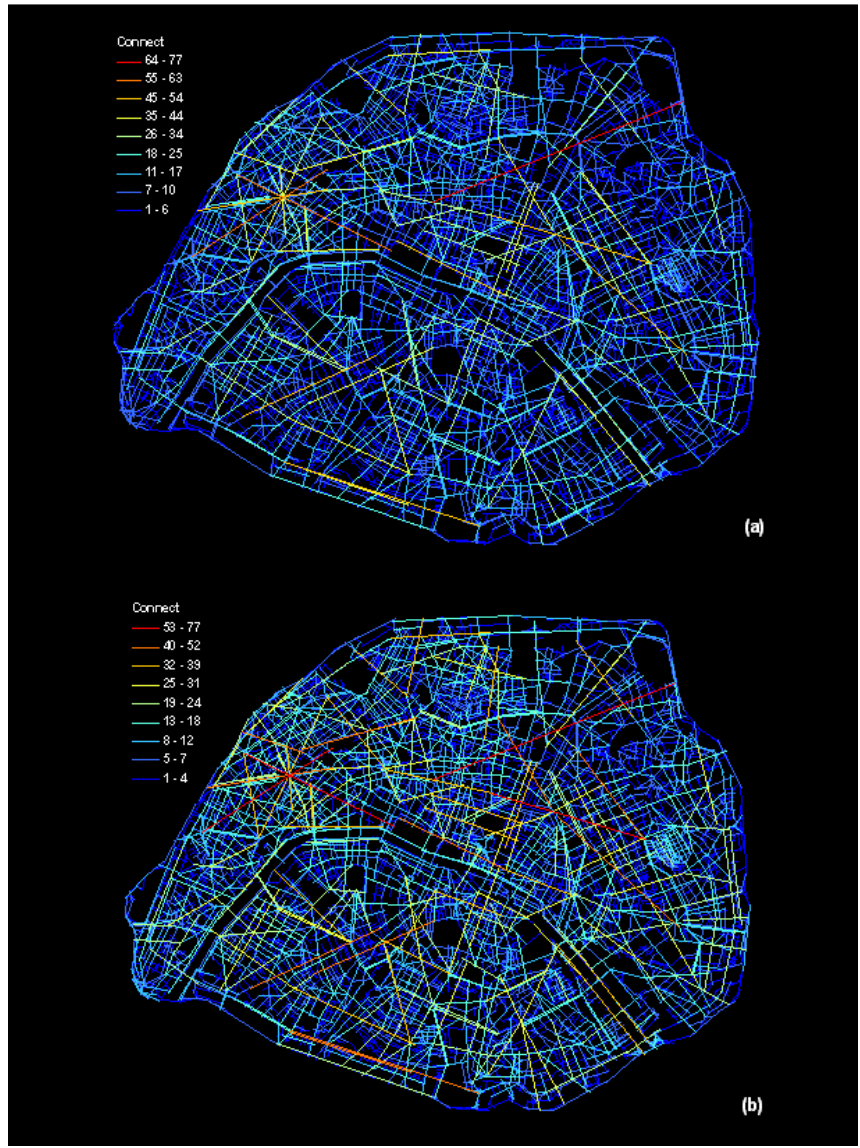

Figure 5: The scaling of Paris streets (in terms of connectivity) captured by the head/tail breaks (a) and the natural breaks (b)

## 5.2 Scaling pattern of Paris streets

The second case study was to map the scaling pattern of Paris streets. The streets are represented by individual axial lines or an axial map, a spatial representation developed for space syntax (Hillier and Hanson 1984). Space syntax provides a set of tools for urban morphological analysis based on free space in urban environments. The axial map consists of the least set of the longest lines cutting across free space. The degree of connectivity of individual lines exhibits a striking long-tailed distribution; there are far more less-connected lines than highly-connected ones. There is a wide range of scales of connectivity for the 6,846 streets, from at least 1 to as many as 77. Using the head/tail breaks, nine hierarchical levels are derived for the Paris streets.

The nine hierarchical levels can be perceived as centers. All lower classes are the ground for upper classes, while upper classes are the figure on the lower classes. Here the terms ground and figure are in terms of figure-ground perception in Gestalt psychology (Rubin 1921). The lower classes tend to reinforce the upper classes to form a whole. It is important to note that the pattern arises from the underlying scaling. Only one street is red, two are orange, and a vast majority were blue. The pattern established through the natural breaks has a similar visual effect (c.f. Figure 5). However, the pattern observed through the head/tail breaks reflected the underlying scaling, while the one represented through the natural breaks did not. This point is striking while examining the class intervals in Table 3. Again, it points to the fact that the head/tail breaks derived classes capture the underlying scaling or hierarchy better than the natural breaks.

Table 3: Nine classes of Paris streets in terms of degree of connectivity

| Classes | 1 | 2 | 3 | 4 | 5 | 6 | 7 | 8 | 9 |
|---|---|---|---|---|---|---|---|---|---|
| Head/tail breaks | 1–6 | 7–10 | 11–17 | 18–25 | 26–34 | 35–44 | 45–54 | 55–63 | 64–77 |
| Number of streets | 5,048 | 1,092 | 483 | 143 | 53 | 18 | 6 | 2 | 1 |
| Natural breaks | 1–4 | 5–7 | 8–12 | 13–18 | 19–24 | 25–31 | 32–39 | 40–52 | 53–77 |
| Number of streets | 3,731 | 1,721 | 897 | 298 | 114 | 47 | 23 | 11 | 4 |

### 5.3 Scaling pattern of population densities

The third case study was to map the population densities of central counties of the state of Kansas, in the United States. Data were taken from the literature (Jenks 1963). Among the 105 areas, the highest population density was 103.4, while the lowest was just 1.6. The data are power-law distributed, as examined in a previous study (Jiang 2012). Using the head/tail breaks, there were four levels of hierarchy or classes, as shown in Figure 6 and Table 4. In perceiving the pattern in Figure 6a, there appeared to be a strong force coming from the lower classes to the upper classes. This is based on the point we made with respect to Figure 2 on how centers reinforce each other. A whole is thus formed in the pattern in which the light colors are the ground, while the dark colors are the figure. The four classes are perceived as four centers, which are organized in a recursive manner. The lighter-color patches support the darker-color patches, and the darkest-color patch is distinguished from many other patches (Figure 6a). This kind of scaling hierarchy is absent in the right side of structure (Figure 6b), in which the two lowest classes occupy almost the same number of values or areas. With the left pattern, the highest level constitutes the strongest center, and there are clearly boundaries between classes or centers. All these factors indicate that the left pattern rather than the right one meet the three structural properties described in Section 3.

Table 4: Four classes of the population densities

| Classes | 1 | 2 | 3 | 4 |
|---|---|---|---|---|
| Head/tail breaks | 1.6–7.6 | 7.6–14.0 | 14.0–45.9 | 45.9–103.4 |
| Number of areas | 74 | 27 | 3 | 1 |
| Natural breaks | 1.6–6.0 | 6.0–18.2 | 18.2–34.6 | 34.6–103.4 |
| Number of areas | 45 | 56 | 2 | 2 |

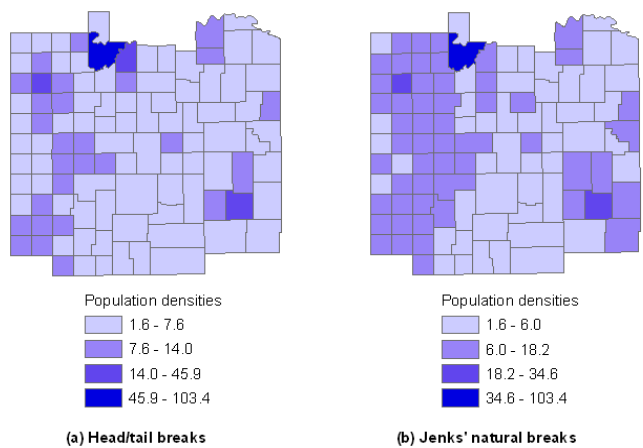

Figure 6: The scaling of the population densities captured by the head/tail breaks (a) rather than the natural breaks (b)

To this point, through reference to Alexander's theory of centers, we have illustrated how beauty can be revealed via the whole or living structure that underlies scaling patterns. As mentioned earlier, beauty is objective and exists at a deep structural level rather than subjective at a superficial level. It may not be instantly or spontaneously perceived. For example, the pattern of Paris streets looks like a haystack, but regularity lurks underneath. To uncover the beauty, one has to conduct the head/tail breaks to derive the hierarchical levels, and further assess the scaling pattern based on the theory of centers to examine how many of the 15 properties are present in the patterns. The case studies deliberately used the same color for the paired patterns to avoid possible biases created by different colors. Based on the theory of centers, all patterns derived from the head/tail breaks should receive a high impact score. The reader should be reminded that the scaling patterns illustrated by symbols or colors with respect to Figure 4, or 5 and 6 arise from the scaling structure. It is the scaling structure rather than symbol or color that evokes the sense of beauty.

## 6. Further discussions on the new kind of beauty in architecture and arts

Previous studies by Salingaros (2005, 2006) and Taylor (2002, 2006) lend further support to our argument that scaling is a beautiful pattern that has positive aesthetic impacts on human beings. Salingaros' work is theoretical in nature, supported by laws and rules from mathematics, physics, and biology. Taylor's work is mostly empirical and inspired by fractal theory. Their conclusions seem to converge with the theme of this paper: the beauty of a scaling pattern lies in its living structure, which consists of many mutually reinforced centers, each of which evokes some degree of beauty or life. Life or beauty is within the structures, substructures, or individual centers. The beauty or life can be quantified or measured objectively (Salingaros 2006), with complex physiological and psychological links to human beings.

A scaling pattern is beautiful because it follows laws of common living structures for physical and biological forms. Inspired by the theory of centers, Salingaros (2006) formulated three laws of the living structure, respectively focusing on small scales, large scales, and the relationship across all scales. In a living structure, scales are not only related to each other, but also link to human emotion and feeling. This linkage gives rise to beauty, being perceived as pleasing, comfortable, and alive. Human beings can directly connect with and touch small scales, so there should be far more such detailed scales. In an architectural space, the smallest scales are mainly for ornamental purposes. In an urban space, the smallest scales are for pedestrians and cycling (for example). On the other hand, the large scales, which humans cannot directly connect with or touch, should be very few and rare. People can reach large scales through intermediate scales. The range of scales from the smallest to the largest constitutes a coherent whole, which has a positive aesthetic appeal or impact.

Related to his third law, and from an architectural design point of view, Salingaros (2006, Chapter 2) proposed a single scaling ratio $x_{n+1}/x_n = 2.7$, the base of the natural logarithm, to form a natural scaling hierarchy. The scaling ratio is based on an early observation that $x_{n+1}/x_n$ should be between 2 and 3 (Alexander 2001). It is well-supported by exponential growth, a fundamental law of nature. This scaling ratio leads to the novel concept of scaling coherence, in which every scale is part of some large scale, and every scale is essential for the scaling hierarchy or whole. The scaling ratio of 2.7 is neither too big nor too small, but just right for connecting different scales and connecting the hierarchy to human consciousness in the sense of emotional satisfaction. This is where harmony or beauty arises. This harmony is essentially different from traditional mathematical harmony such as the golden ratio. The golden ratio is part of Euclidean geometry, while scaling coherence belongs to fractal mathematics and complexity theories.

Fractals are perceived to be pleasing and have positive aesthetic impacts on human well-being. Fractals or scaling can be understood with reference to the simple truism that there are "far more small things than large ones", although fractal dimension is more in the sense of space filling. Fractals were developed in the 1960s by Mandelbrot (1967), but many mathematical fractals such as the Koch snowflake were recognized and studied far earlier. All of these mathematical fractals have self-similar or scaling coherent properties. Mandelbrot's major finding was statistical self-similarity or statistical scaling, observed in many natural, economic, and societal phenomena. It is believed that fractals might be biological, or that human minds are hard-wired to perceive fractals as pleasing and comfortable. For example, a natural scene through a window has significant positive impacts on a hospital patient's recovery from surgery (Ulrich 1984). The essence of the natural scene that has positive impact on the patient's recovery is fractals or scaling.

Even artificially created scaling scenes or computer-generated fractals have the same effect on human well-being. The previous work by Ulrich (1984), Taylor (2006) and Alexander (1993, 2001) leads us to believe that it is the underlying scaling structure that triggers the perception or experience of beauty. In other words, if the patient stayed in a room that was decorated with computer-generated fractal patterns, it would have the same positive effect as the natural scene through a window. In this regard, it is worth noting that Taylor's major contribution is his discovery of fractals within Pollock's poured paintings (Taylor 2002). The poured paintings of Jackson Pollock, an influential American painter, triggered considerable controversy. Some people celebrated his works, while others categorically disliked and dismissed them. For those who dislike the paintings, the main reason is their abstract expressionistic style. It is hard for an observer to see anything meaningful from the paintings in an intuitive sense. Taylor (2002) systematically studied Pollock's poured paintings using fractal mathematics, and measured fractal dimensions to characterize Pollock's painting at different time periods. It was found that the fractal dimension decreased with time. Taylor also used fractal dimensions as a criterion with which to rank the quality of the paintings; he suggested based on his empirical studies that a fractal dimension of 1.3 to 1.5 has the highest positive impacts on human well-being.

Pollock painted his fractal-like paintings more than 20 years before the scientific community discovered fractals, so it is unlikely that the paintings were consciously guided by the concept. Instead, his painting skills were developed unconsciously, and driven by something deep in his unconscious or psyche. Pollock himself admitted that he wanted to deal with the imagery lurking in the dark depths of his unconscious. He was not interested in mimicking nature, but his work nonetheless expressed the essence of nature that is scaling. In other words, Pollack captured the essence rather than the appearance of nature. His famous painting "Blue Poles" was done in a drunken, suicidal state on a stormy night in March 1952. This indicates the fact that his unconscious played an important role in the painting process. We human beings might be hardwired to construct or perceive fractal or scaling things.

## 7. Conclusion

This paper attempts to argue that a new kind of beauty arises from the underlying scaling of geographic space. This argument is well supported by the theory of centers, and related studies in architecture and arts. The theory of centers provides a set of concepts to objectively assess or appreciate the beauty lurking underneath structures. This is the beauty identified through objective judgment, rather than as a matter of personal preference or taste; the nature of beautiful order is not just psychological, but physical (Alexander 2001). This may sound provocative to many readers. Based on the theory and case studies, we have illustrated that the scaling patterns of geographic space can indeed evoke a sense of beauty founded on a very intense field of centers. We have shown that the head/tail breaks rather than the natural breaks can capture the underlying scaling. Our argument echoes the famous line - *"beauty is truth, truth beauty."* Here, the truth is the underlying scaling structure, which is universal and widely seen in physical, biological, and built forms, as well as in good arts and designs. This is what underlies the new kind of science behind *The Nature of Order*.

The theory of centers has been further articulated and developed by Salingaros (2005, 2006) to better design architectural space and urban environments, and to appreciate the beauty (or note the ugliness) of our surroundings at a deep structural level. This is essentially different from traditional definition of beauty. For example, the ancient Greeks traditionally believed that beauty was constituted by three ingredients: symmetry, proportion, and harmony. In fact, scaling is disproportional with respect to the imbalance between the heads and the tails. The scaling patterns may look rather disordered on the surface, but the order is underneath and universal. One needs to expend effort to uncover the beauty, as shown in the case studies. The more one looks at the pattern, the more a sense of aesthetic feeling emerges. These aesthetic rules apply whether designing a house, weaving a carpet, or designing a map. Future work will seek to quantify the beauty via mathematics, following Alexander in his call for effective mathematics that would measure the degree of life of the centers or the whole.